\documentclass{PoS}

\title{Next-to-leading order multi-leg processes for the Large Hadron Collider}

\ShortTitle{NLO processes for LHC}

\author{\speaker{Thomas Binoth}, Thomas Reiter\thanks{This work was supported by the British Science and Technology Facilities Council (STFC), the Scottish Universities Physics Alliance (SUPA)
and the Deutsche Forschungsgemeinschaft (DFG, BI-1050/1).}\\
        School of Physics, The University of Edinburgh, Edinburgh EH9 3JZ, UK.\\
        E-mail: \email{binoth@ph.ed.ac.uk,thomas.reiter@ed.ac.uk}}

\author{Jeppe R. Andersen\\
        CERN, CH-1211 Geneva, Switzerland.\\
        E-mail: \email{jeppe.andersen@cern.ch}}

\author{Gudrun Heinrich, Jennifer M. Smillie\\
        IPPP, University of Durham, Durham DH1 3LE, UK.\\
        E-mail: \email{gudrun.heinrich@durham.ac.uk,j.m.smillie@durham.ac.uk}}

\author{Jean-Philippe Guillet, Gregory Sanguinetti \\
        LAPTH, 9, Chemin de Bellevue BP 110, 74941 Annecy le Vieux, France.\\
        E-mail: \email{guillet@lapp.in2p3.fr,sangui@lapp.in2p3.fr}}

\author{Stefan Karg\\
        Institute for Theoretical Physics E, RWTH Aachen, D-52056 Aachen, Germany.\\
        E-mail: \email{karg@physik.rwth-aachen.de}}

\author{Nikolas Kauer\\
        Institute for Theoretical Physics, University of W\"urzburg, 
	D-97074 W\"urzburg, Germany.\\
        E-mail: \email{kauer@physik.uni-wuerzburg.de}}

\abstract{In this talk we discuss recent progress concerning precise predictions
for the LHC. We give a status report of three applications of our method to deal 
with multi-leg one-loop amplitudes: The interference term of Higgs production by
gluon- and weak boson fusion to order $\mathcal{O}(\alpha^2\alpha_s^3)$ and the next-to-leading order corrections
to the two processes $pp\to ZZ\,\mbox{jet}$ and $u\bar{u}\to d\bar{d}s\bar{s}$. The latter is a subprocess of the 
four jet cross section at the LHC.}

\FullConference{8th International Symposium on Radiative Corrections (RADCOR)\\
		 October 1-5 2007\\
		 Florence, Italy}

\begin{document}

\section{Introduction}

The Large Hadron Collider (LHC) at CERN will probe our understanding of electroweak symmetry
breaking and explore physics in the TeV region. A detailed theoretical knowledge of various kinds
of Standard Model backgrounds is indispensable for these studies.
Especially in the startup-phase when it is necessary to calibrate detectors
 using multi-jet signatures, 
preferably with identifiable leptons, Standard Model processes will play 
an important role.  Precise predictions for
such multi-partonic cross sections are only possible by including
higher order corrections such that renormalisation and factorisation scale
dependencies are tamed. While corrections at next-to-leading order  
in the strong coupling constant $\alpha_s$ are known
to basically all relevant $2\to 2$ processes,  the situation for $2\to N$ processes 
whith $N\geq 3$ is less satisfactory, although tremendous progress has been
made in the last few years. For $2\to 3$ processes, various methods have been used
recently to obtain NLO QCD predictions for multi-boson production $pp\to ZZZ,WWZ,HHH$ 
\cite{Lazopoulos:2007ix,Hankele:2007sb,Plehn:2005nk,Binoth:2006ym},
processes in the context of weak boson fusion, like $pp\to WWjj,WZjj$ \cite{Jager:2006zc,Bozzi:2007ur}, 
$pp\to Hjj$ with  effective gluon-Higgs
couplings \cite{Campbell:2006xx},
$gg\to Hq\bar{q}$ \cite{Weber:2006au}, and $pp\to t\bar{t}j$ \cite{Dittmaier:2007wz}. 

Many different techniques are used for the evaluation of multi-particle 
processes, as a result of recent developments which were  
triggered by the observation that 
the standard Passarino-Veltman reduction in general does not
lead to numerically stable amplitude representations for the reduction
of five-point integrals.
Apart from Feynman diagrammatic approaches which apply new reduction
techniques for some or all scalar and tensor integrals 
\cite{Giele:2004iy,Binoth:2005ff,Denner:2005nn}
the evaluation of one loop amplitudes by 
extracting the coefficients of a certain basis set of scalar 
integrals  using unitarity based methods~\cite{Bern:2007dw} both in  algebraic 
\cite{Britto:2006sj,Binoth:2007ca,Bernicot:2007hs} 
and numerical\cite{Ossola:2006us,Ossola:2007bb,Ellis:2007br} variants has seen 
substantial progress recently. Purely numerical approaches based on 
Feynman diagrams, 
which do not use any reduction to ``basis integrals'' are also 
viable~\cite{Lazopoulos:2007ix,Anastasiou:2007qb,Nagy:2006xy,Ferroglia:2002mz}.

In \cite{Binoth:2005ff} we have proposed a framework for the evaluation of one-loop
multi-leg amplitudes, based on reduction formulas in Feynman parameter 
space~\cite{Bern:1993kr,Binoth:1999sp} 
in the context of dimensional regularisation.
The reduction of rank $R$ $N$-point to rank $R$-1 ($N$-1)-point integrals can be obtained
algebraically for $N>5$. For $N\leq 5$ we provide form factor representations
which are expressed in terms of  3- and 4-point scalar integrals with
Feynman parameters in the numerator.
The dimensionality of the box functions is such that all IR divergences,
i.e. poles in $1/(n-4)$, are isolated into the triangle functions.
For all  IR divergent triangle functions explicit representations
can be obtained.  We have coded all formulas up to $N=6$ for massless internal kinematics
into a \texttt{FORTRAN90} code, called \texttt{golem90} \cite{inprep}.  
The code allows to switch between a semi-numerical and completely numerical evaluation of the
basis functions. The latter is preferable in exceptional phase space regions where
certain basis integrals can become linearly dependent.
In this way the  problems of instabilities due to inverse Gram determinants
can be tamed. 
Optionally, our reduction formalism also allows a reduction to a
scalar integral basis of 1-, 2-, 3-point functions in 
$n$ dimensions  and 
4-point functions in $n$+2 dimensions, denoted by $I_1^n$, $I_2^n$, $I_3^n$ $I_4^{n+2}$. 
This fully algebraic reduction is to be used away from exceptional phase space points, 
where it is fast and reliable.

In the following  we will discuss three applications of our 
method relevant for LHC phenomenology.

\section{Interference term for $pp\to Hjj$ at order $\mathcal{O}(\alpha^2\alpha_s^3)$}

Weak boson fusion is one of the most promising discovery channels for the Higgs
boson. As such it deserves a careful consideration of higher order effects. 
Very recently, the electroweak  corrections, including a recalculation of QCD corrections
have been evaluated in \cite{Ciccolini:2007ec}.
A Born level interference term between the gluon fusion and weak boson fusion 
is only allowed by colour conservation if the in- and outgoing quarks are crossed
in the  $t\leftrightarrow u$-channel, 
which is kinematically disfavoured. 
Such interference terms are included in the 
calculations of  \cite{Andersen:2006ag,Ciccolini:2007ec}.
However, the exchange of an extra gluon between the quark lines
opens up a viable colour channel. 
We have obtained a fully analytic result of this one-loop interference term 
between gluon fusion and weak boson fusion~\cite{Andersen:2007mp}, and
implemented the evaluation in a flexible \texttt{C++} Monte Carlo programme.

\begin{figure}[tbp]
  \centering
  \includegraphics[width=0.48\textwidth]{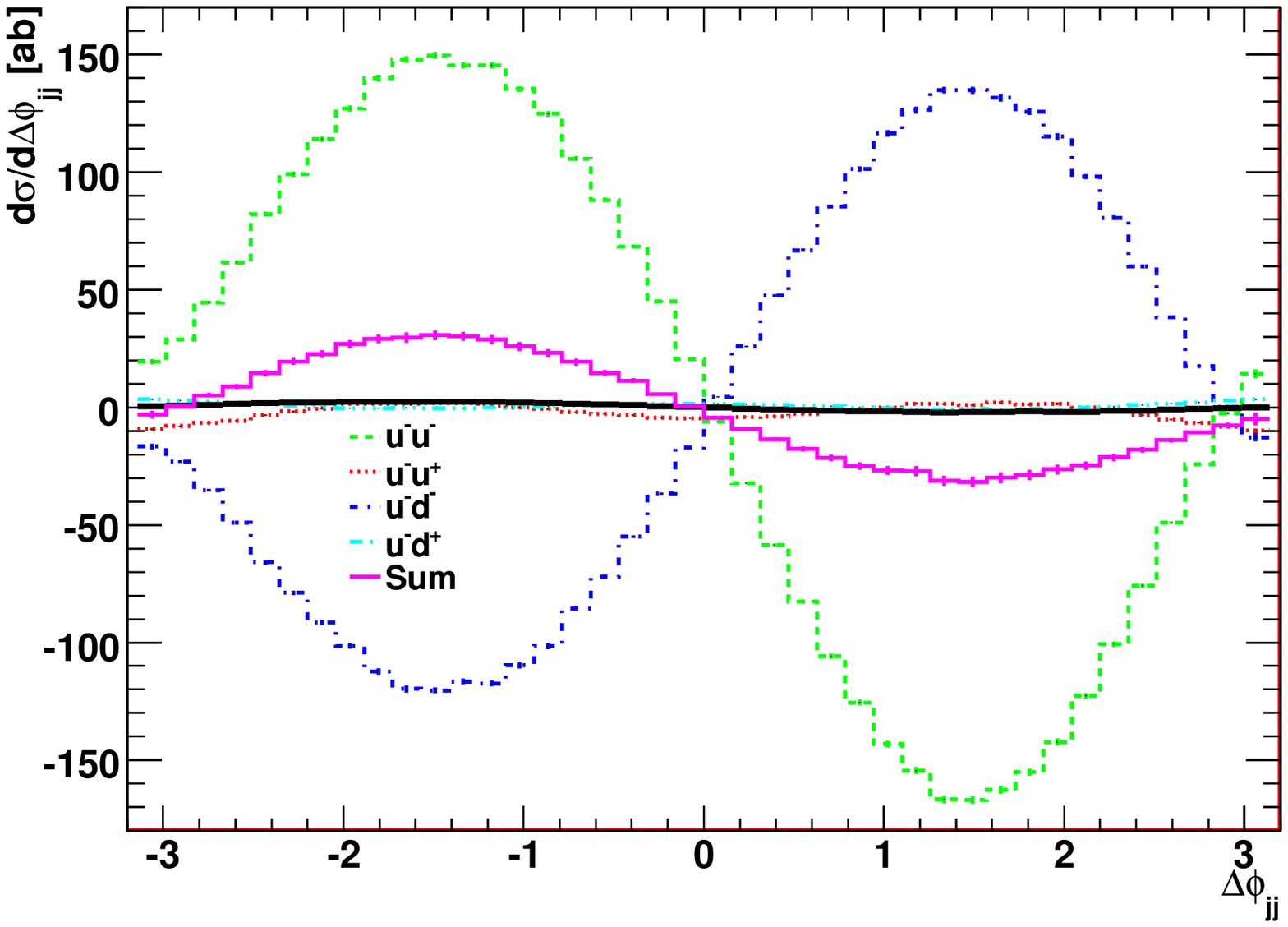}
  \includegraphics[width=0.48\textwidth]{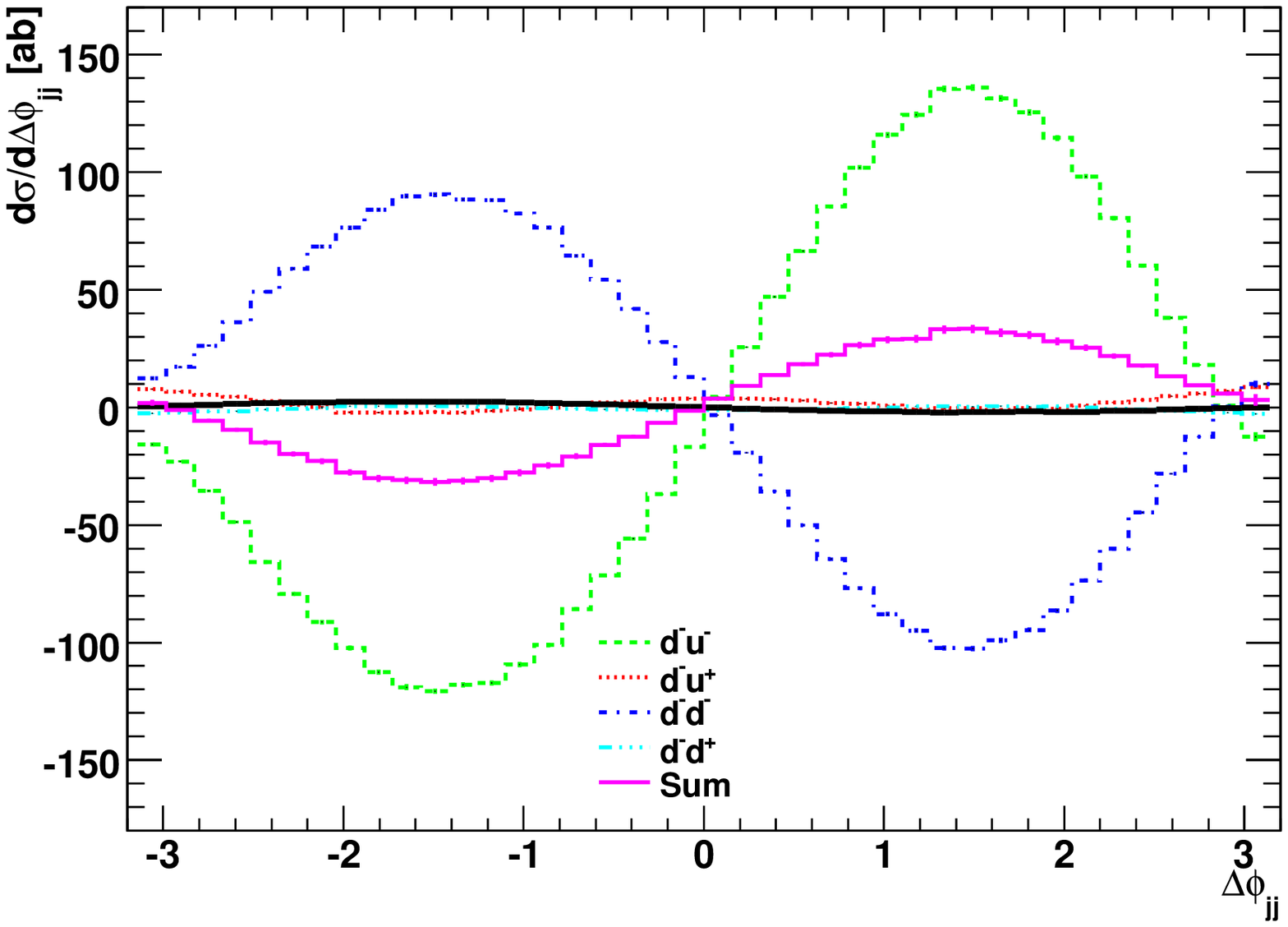}
  \caption{The $\Delta\phi_{jj}$-distribution for various flavour and
    helicity-configurations. The purple histogram labelled ``Sum'' indicates
    the sum over the four contributions shown. The sum over \emph{all}
    flavour and helicity assignments including all sea flavours is shown in the 
    black histogram. WBF cuts have been used \cite{Andersen:2007mp}.}
  \label{fig:flavhel_seaval}
\end{figure}

Figure~\ref{fig:flavhel_seaval} displays the contribution to the distribution in
$\Delta\phi_{jj}$ from the interference terms for various helicity and flavour
configurations  for a Higgs boson mass of 115~GeV.

Note that the integral of the absolute
value of the $\Delta\phi_{jj}$ distribution, 
$$\int_{-\pi}^\pi d \Delta\phi_{jj} |\frac{d\sigma}{d\Delta\phi_{jj}}|\;\;,$$
is a useful  measure of the impact of the
interference effect on the extraction of the $ZZH$-vertex. 
This integral evaluates  to $9.1$~ab.
%, an order of magnitude larger than the integral over the oscillating distribution.
The total integral over the absolute value of the fully differential 
cross section leads  to $29.6$~ab. We conclude that
the interference term can be safely neglected in phenomenological studies, 
but this  needed to be checked by doing the explicit calculation.

\section{The virtual $\mathcal{O}(\alpha_s)$ corrections to $pp\to ZZj$}

During the Les Houches 2005 workshop the process
$pp \to VVj$ was identified as one of the most important
missing NLO calculations \cite{Buttar:2006zd}. The process with a charged  vector boson pair 
has been evaluated very recently by two independent groups \cite{Dittmaier:2007th,Campbell:2007ev}. 
The evaluation for ZZ plus jet is still missing. 
The process is composed  of three partonic reactions
\begin{equation}
q\bar{q} \to ZZg \; , \; gq \to ZZq \; , \; g\bar{q} \to ZZ\bar{q}\; , \nonumber
\end{equation}
With our methods we have obtained the virtual  order $\mathcal{O}(\alpha_s)$ corrections for 
all  helicity amplitudes of both processes.  
Using spinor helicity methods we have obtained analytical formulas
for the coefficients of all basis scalar integrals.
We work in dimensional regularisation and treat $\gamma_5$ by applying
the 't~Hooft-Veltman scheme. As an illustration we show
the contribution of the virtual correction to some typical 
distributions. Only the contributions which are related to finite
basis integrals are plotted. 
 \begin{figure}[tbp]
  \centering
  \includegraphics[width=0.48\textwidth]{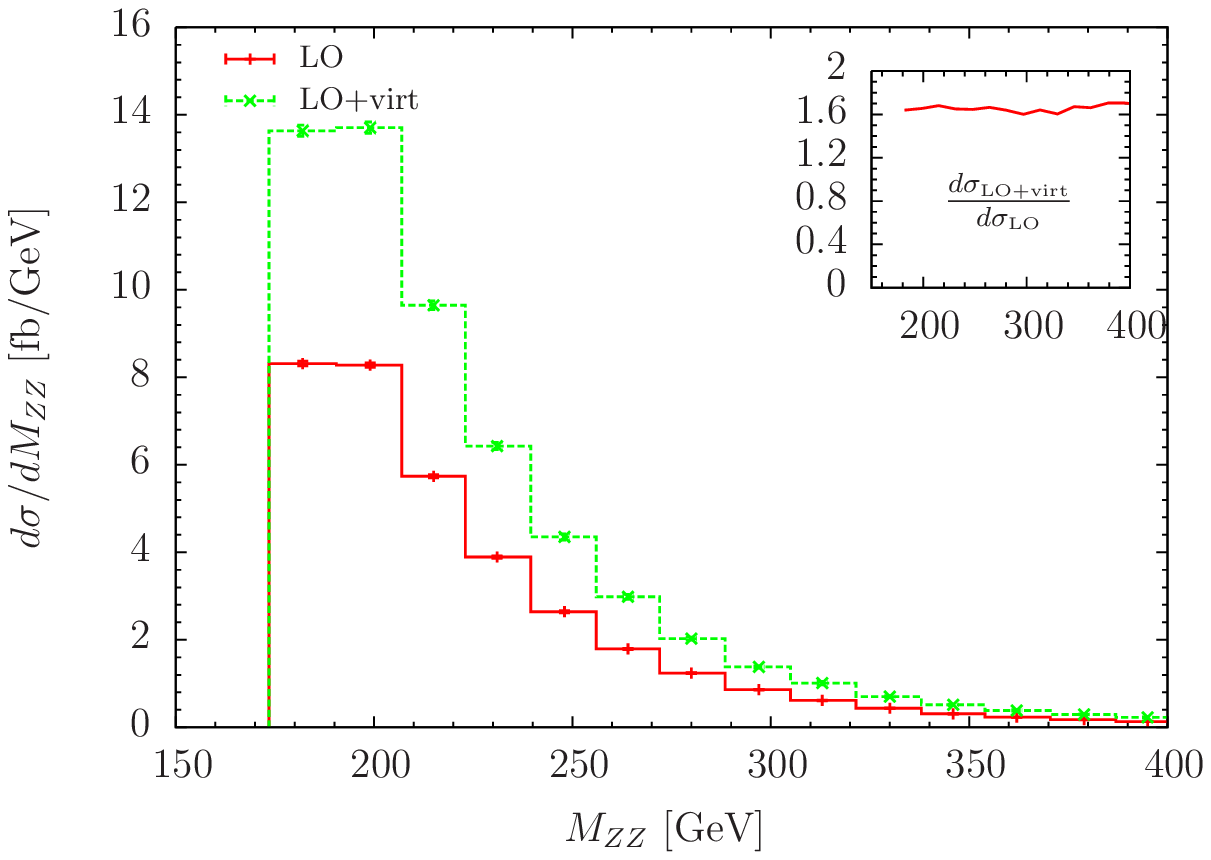}
  \includegraphics[width=0.48\textwidth]{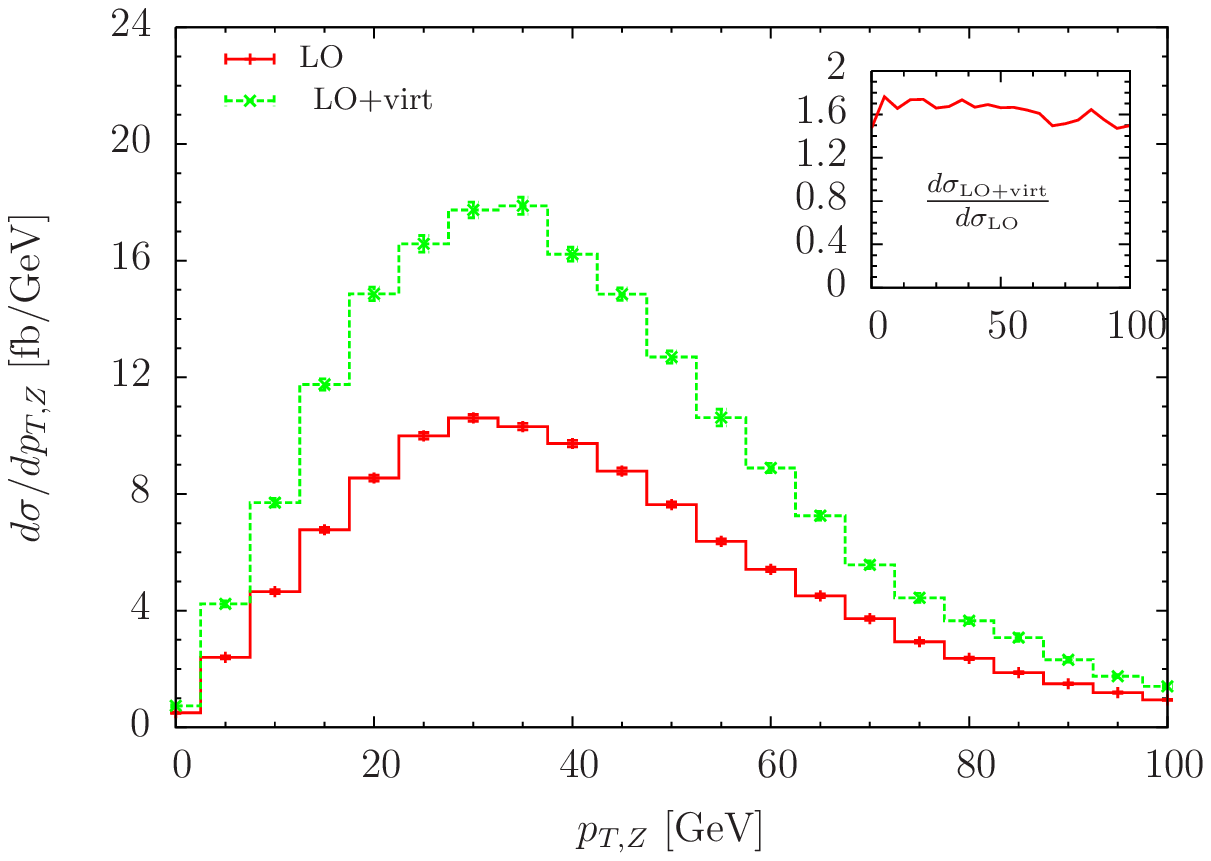}
  \caption{The finite virtual NLO contribution to the helicity component $--+++$ of the
           partonic process $q\bar{q}\to ZZg$. The invariant mass of the Z pair is 
	   shown on the left
	   and the $p_T$ distribution on the right. We use the cut 
	   $p_{T,\,\textrm{jet}} > 20$ GeV
           and a parton and beam pipe separation cut of  $\theta_{ij} > 1.5^0$.}
  \label{fig:qqbarzzg}
\end{figure}
For the full result the real emission corrections remain to be included \cite{inprep1}.

\section{The amplitude  $u\bar{u}\to d\bar{d}s\bar{s}$}

Not a single NLO $2\to 4$ process relevant for LHC phenomenology has been evaluated so far.
As a test case for our reduction methods we have evaluated the 6-photon 
amplitude \cite{Binoth:2007ca} and have compared our result with an evaluation using a fully
numerical approach \cite{Nagy:2006xy} and unitarity
based methods \cite{Ossola:2007bb,Bernicot:2007hs}.  
The same set-up can be used to attack now processes like $pp\to jjjj$
and $pp\to b\bar{b}b\bar{b}$ which are of relevance for background studies at the LHC.
As an example we show the result of one colour factor of the finite 
virtual NLO contribution to the $u\bar{u}\to d\bar{d}s\bar{s}$ amplitude
in massless QCD.
The calculation has been carried out using spinor helicity
amplitudes in the 't~Hooft-Veltman scheme.
We have chosen a convenient colour basis,
which allows to split the amplitude as follows
\begin{equation}
\sum_{\lambda}
\sum_{i=1}^6
{\mathbf C}^iA^{\lambda}_i(p_1,\ldots,p_6),
\end{equation}
where $A_c^{\lambda}$ are the helicity and colour sub-amplitudes. In
particular we chose the colour structures
\begin{equation}
\vec{{\mathbf C}}=
(\delta_{c_1}^{c_2}\delta_{c_4}^{c_3}\delta_{c_6}^{c_5},
\delta_{c_1}^{c_2}\delta_{c_4}^{c_5}\delta_{c_6}^{c_3},
\delta_{c_1}^{c_5}\delta_{c_4}^{c_2}\delta_{c_6}^{c_3},
\delta_{c_1}^{c_5}\delta_{c_4}^{c_3}\delta_{c_6}^{c_2},
\delta_{c_1}^{c_3}\delta_{c_4}^{c_5}\delta_{c_6}^{c_2},
\delta_{c_1}^{c_3}\delta_{c_4}^{c_2}\delta_{c_6}^{c_5}).
\end{equation}
In our notation $\lambda$ is the vector $(\lambda_1,\ldots,\lambda_6)$,
and $\lambda_j=\pm1$ is the helicity of the particle with momentum $p_j$
of which the colour index is $c_j$. In the six-quark amplitude one can
identify two independent helicity configurations, $\lambda^a=(+,+,+,+,+,+)$ and
$\lambda^b=(+,+,+,+,-,-)$.

\begin{figure}
\begin{center}
\includegraphics[width=0.5\textwidth]{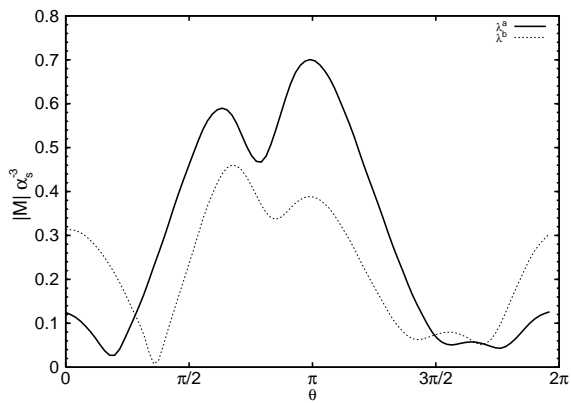}
\caption{Six-quark amplitude. The finite parts  of
the two helicity configurations $s\vert A_1^{\lambda^a}\vert\alpha_s^{-3}$ (solid)
and $s\vert A_1^{\lambda^b}\vert\alpha_s^{-3}$ (dashed)
are plotted for an arbitrary
kinematical point where the final state momenta have been rotated about
the $y$-axis by an angle $\theta$.}
\label{fig:sixquarks:rotplot;golem}
\end{center}
\end{figure}

We  reduced the tensor integrals
to form factors as outlined above (for more details see~\cite{Binoth:2005ff}), 
and  deal with the spinor algebra by
completing spinor lines to traces.
The expressions for the
diagrams are transformed into a \texttt{Fortran90} program. The \texttt{golem90}
library is used for the numerical evaluation of the form factors.
The code returns the sub-amplitudes in the form
\begin{equation}
A^\lambda_i(p_1,\ldots,p_6)=\frac{g_s^6}{4\pi^2}\frac{1}{s}%
\left(\frac{A}{\varepsilon^2}+\frac{B}{\varepsilon}+C
+{\mathcal O}(\varepsilon)\right)
\end{equation}
for each of the six colour structures and for all non-zero helicities,
where $A$, $B$ and $C$ are complex coefficients. As an example 
in Fig.~\ref{fig:sixquarks:rotplot;golem} we plot
the amplitude $s\vert A_c^\lambda\vert\alpha_s^{-3}$ for the colour structure
$c=1$ and the two helicity configurations $\lambda^a$ and $\lambda^b$. 
The initial state momenta
have been fixed to be aligned with the $z$-axis while the final state momenta have been
rotated about the $y$-axis by an angle $\theta$. For $\theta=0$ the momenta
are chosen as in Ref.~\cite{Nagy:2006xy}.
In the chosen units the renormalisation scale is $\mu=1$.
The amplitude has been evaluated at 50 successive points between
$\theta=0$ and $\theta=2\pi$, which took
2.4~seconds per point and helicity on an Intel Pentium~4 CPU (3.2\,GHz).

\section{Conclusion}

The LHC demands next-to-leading order precision for many
multi-particle processes. We have presented in this talk 
some recent results using our one-loop reduction methods, 
ranging from completely massless amplitudes to 
ones with several mass scales. 
The progress which has been made in the last years
in the context of precision phenomenology 
by many groups is good news for the LHC. Many relevant
predictions beyond leading order are or will become available 
in the near future and there is a clear movement towards largely automated 
NLO tools as well as combining NLO amplitudes with parton showers.


\begin{thebibliography}{99}


%\cite{Lazopoulos:2007ix}
\bibitem{Lazopoulos:2007ix}
  A.~Lazopoulos, K.~Melnikov and F.~Petriello,
  %``QCD corrections to tri-boson production,''
  Phys.\ Rev.\  D {\bf 76} (2007) 014001
  [arXiv:hep-ph/0703273].
  %%CITATION = PHRVA,D76,014001;%%

%\cite{Hankele:2007sb}
\bibitem{Hankele:2007sb}
  V.~Hankele and D.~Zeppenfeld,
  %``QCD corrections to hadronic WWZ production with leptonic decays,''
  arXiv:0712.3544 [hep-ph].
  %%CITATION = ARXIV:0712.3544;%%

%\cite{Plehn:2005nk}
\bibitem{Plehn:2005nk}
  T.~Plehn and M.~Rauch,
  %``The quartic Higgs coupling at hadron colliders,''
  Phys.\ Rev.\  D {\bf 72} (2005) 053008
  [arXiv:hep-ph/0507321].
  %%CITATION = PHRVA,D72,053008;%%


%\cite{Binoth:2006ym}
\bibitem{Binoth:2006ym}
  T.~Binoth, S.~Karg, N.~Kauer and R.~Ruckl,
  %``Multi-Higgs boson production in the standard model and beyond,''
  Phys.\ Rev.\  D {\bf 74} (2006) 113008
  [arXiv:hep-ph/0608057].
  %%CITATION = PHRVA,D74,113008;%%

%\cite{Lazopoulos:2007bv}
%\bibitem{Lazopoulos:2007bv}
%  A.~Lazopoulos, K.~Melnikov and F.~J.~Petriello,
  %``NLO QCD corrections to the production of t-tbar-Z in gluon fusion,''
%  arXiv:0709.4044 [hep-ph].
  %%CITATION = ARXIV:0709.4044;%%

%\cite{Jager:2006zc}
\bibitem{Jager:2006zc}
  B.~Jager, C.~Oleari and D.~Zeppenfeld,
  %``Next-to-leading order QCD corrections to W+ W- production via vector-boson
  %fusion,''
  JHEP {\bf 0607} (2006) 015
  [arXiv:hep-ph/0603177].
  %%CITATION = JHEPA,0607,015;%%

%\cite{Bozzi:2007ur}
\bibitem{Bozzi:2007ur}
  G.~Bozzi, B.~Jager, C.~Oleari and D.~Zeppenfeld,
  %``Next-to-leading order QCD corrections to W+Z and W-Z production via
  %vector-boson fusion,''
  Phys.\ Rev.\  D {\bf 75} (2007) 073004
  [arXiv:hep-ph/0701105].
  %%CITATION = PHRVA,D75,073004;%%

%\cite{Campbell:2006xx}
\bibitem{Campbell:2006xx}
  J.~M.~Campbell, R.~K.~Ellis and G.~Zanderighi,
  %``Next-to-leading order Higgs + 2 jet production via gluon fusion,''
  JHEP {\bf 0610} (2006) 028
  [arXiv:hep-ph/0608194].
  %%CITATION = JHEPA,0610,028;%%

%\cite{Weber:2006au}
\bibitem{Weber:2006au}
  M.~M.~Weber,
  %``Gluon initiated vector boson fusion,''
  Nucl.\ Phys.\ Proc.\ Suppl.\  {\bf 160} (2006) 200.
  %%CITATION = NUPHZ,160,200;%%

%\cite{Dittmaier:2007wz}
\bibitem{Dittmaier:2007wz}
  S.~Dittmaier, P.~Uwer and S.~Weinzierl,
  %``NLO QCD corrections to t anti-t + jet production at hadron colliders,''
  Phys.\ Rev.\ Lett.\  {\bf 98} (2007) 262002
  [arXiv:hep-ph/0703120].
  %%CITATION = PRLTA,98,262002;%%

%\cite{Giele:2004iy}
\bibitem{Giele:2004iy}
  W.~T.~Giele and E.~W.~N.~Glover,
  %``A calculational formalism for one-loop integrals,''
  JHEP {\bf 0404} (2004) 029
  [arXiv:hep-ph/0402152].
  %%CITATION = JHEPA,0404,029;%%

%\cite{Binoth:2005ff}
\bibitem{Binoth:2005ff}
  T.~Binoth, J.~P.~Guillet, G.~Heinrich, E.~Pilon and C.~Schubert,
  %``An algebraic / numerical formalism for one-loop multi-leg amplitudes,''
  JHEP {\bf 0510} (2005) 015
  [arXiv:hep-ph/0504267].
  %%CITATION = JHEPA,0510,015;%%

%\cite{Denner:2005nn}
\bibitem{Denner:2005nn}
  A.~Denner and S.~Dittmaier,
  %``Reduction schemes for one-loop tensor integrals,''
  Nucl.\ Phys.\  B {\bf 734} (2006) 62
  [arXiv:hep-ph/0509141].
  %%CITATION = NUPHA,B734,62;%%

%\cite{Bern:2007dw}
\bibitem{Bern:2007dw}
  Z.~Bern, L.~J.~Dixon and D.~A.~Kosower,
  %``On-Shell Methods in Perturbative QCD,''
  Annals Phys.\  {\bf 322} (2007) 1587
  [arXiv:0704.2798 [hep-ph]].
  %%CITATION = APNYA,322,1587;%%


%\cite{Britto:2006sj}
\bibitem{Britto:2006sj}
  R.~Britto, B.~Feng and P.~Mastrolia,
  %``The cut-constructible part of QCD amplitudes,''
  Phys.\ Rev.\  D {\bf 73} (2006) 105004
  [arXiv:hep-ph/0602178].
  %%CITATION = PHRVA,D73,105004;%%

%\cite{Binoth:2007ca}
\bibitem{Binoth:2007ca}
  T.~Binoth, G.~Heinrich, T.~Gehrmann and P.~Mastrolia,
  %``Six-Photon Amplitudes,''
  Phys.\ Lett.\  B {\bf 649} (2007) 422
  [arXiv:hep-ph/0703311].
  %%CITATION = PHLTA,B649,422;%%

%\cite{Bernicot:2007hs}
\bibitem{Bernicot:2007hs}
  C.~Bernicot and J.~P.~Guillet,
  %``Six-Photon Amplitudes in Scalar QED,''
  arXiv:0711.4713 [hep-ph].
  %%CITATION = ARXIV:0711.4713;%%

%\cite{Ossola:2006us}
\bibitem{Ossola:2006us}
  G.~Ossola, C.~G.~Papadopoulos and R.~Pittau,
  %``Reducing full one-loop amplitudes to scalar integrals at the integrand
  %level,''
  Nucl.\ Phys.\  B {\bf 763} (2007) 147
  [arXiv:hep-ph/0609007].
  %%CITATION = NUPHA,B763,147;%%

%\cite{Ossola:2007bb}
\bibitem{Ossola:2007bb}
  G.~Ossola, C.~G.~Papadopoulos and R.~Pittau,
  %``Numerical Evaluation of Six-Photon Amplitudes,''
  JHEP {\bf 0707} (2007) 085
  [arXiv:0704.1271 [hep-ph]].
  %%CITATION = JHEPA,0707,085;%%


  %\cite{Ellis:2007br}
\bibitem{Ellis:2007br}
  R.~K.~Ellis, W.~T.~Giele and Z.~Kunszt,
  %``A Numerical Unitarity Formalism for Evaluating One-Loop Amplitudes,''
  arXiv:0708.2398 [hep-ph].
  %%CITATION = ARXIV:0708.2398;%%

%\cite{Anastasiou:2007qb}
\bibitem{Anastasiou:2007qb}
  C.~Anastasiou, S.~Beerli and A.~Daleo,
  %``Evaluating multi-loop Feynman diagrams with infrared and threshold
  %singularities numerically,''
  JHEP {\bf 0705} (2007) 071
  [arXiv:hep-ph/0703282].
  %%CITATION = JHEPA,0705,071;%%

%\cite{Nagy:2006xy}
\bibitem{Nagy:2006xy}
  Z.~Nagy and D.~E.~Soper,
  %``Numerical integration of one-loop Feynman diagrams for N-photon
  %amplitudes,''
  Phys.\ Rev.\  D {\bf 74} (2006) 093006
  [arXiv:hep-ph/0610028].
  %%CITATION = PHRVA,D74,093006;%%

%\cite{Ferroglia:2002mz}
\bibitem{Ferroglia:2002mz}
  A.~Ferroglia, M.~Passera, G.~Passarino and S.~Uccirati,
  %``All-purpose numerical evaluation of one-loop multi-leg Feynman diagrams,''
  Nucl.\ Phys.\  B {\bf 650} (2003) 162
  [arXiv:hep-ph/0209219].
  %%CITATION = NUPHA,B650,162;%%
  
%\cite{Bern:1993kr}
\bibitem{Bern:1993kr}
  Z.~Bern, L.~J.~Dixon and D.~A.~Kosower,
  %``Dimensionally regulated pentagon integrals,''
  Nucl.\ Phys.\  B {\bf 412} (1994) 751
  [arXiv:hep-ph/9306240].
  %%CITATION = NUPHA,B412,751;%%

%\cite{Binoth:1999sp}
\bibitem{Binoth:1999sp}
  T.~Binoth, J.~P.~Guillet and G.~Heinrich,
  %``Reduction formalism for dimensionally regulated one-loop N-point
  %integrals,''
  Nucl.\ Phys.\  B {\bf 572} (2000) 361
  [arXiv:hep-ph/9911342].
  %%CITATION = NUPHA,B572,361;%%

%\cite{Ciccolini:2007ec}
\bibitem{Ciccolini:2007ec}
  M.~Ciccolini, A.~Denner and S.~Dittmaier,
  %``Electroweak and QCD corrections to Higgs production via vector-boson fusion
  %at the LHC,''
  arXiv:0710.4749 [hep-ph].
  %%CITATION = ARXIV:0710.4749;%%

%\cite{Andersen:2006ag}
\bibitem{Andersen:2006ag}
  J.~R.~Andersen and J.~M.~Smillie,
  %``QCD and electroweak interference in Higgs production by gauge boson
  %fusion,''
  Phys.\ Rev.\  D {\bf 75}, 037301 (2007)
  [arXiv:hep-ph/0611281].
  %%CITATION = PHRVA,D75,037301;%%

%\cite{Andersen:2007mp}
\bibitem{Andersen:2007mp}
  J.~R.~Andersen, T.~Binoth, G.~Heinrich and J.~M.~Smillie,
  %``Loop induced interference effects in Higgs Boson plus two jet production at
  %the LHC,''
  arXiv:0709.3513 [hep-ph].
  %%CITATION = ARXIV:0709.3513;%%

%\cite{Buttar:2006zd}
\bibitem{Buttar:2006zd}
  C.~Buttar {\it et al.},
  %``Les Houches physics at TeV colliders 2005, standard model, QCD, EW, and
  %Higgs working group: Summary report,''
  arXiv:hep-ph/0604120.
  %%CITATION = HEP-PH/0604120;%%

%\cite{Dittmaier:2007th}
\bibitem{Dittmaier:2007th}
  S.~Dittmaier, S.~Kallweit and P.~Uwer,
  %``NLO QCD corrections to WW+jet production at hadron colliders,''
  arXiv:0710.1577 [hep-ph].
  %%CITATION = ARXIV:0710.1577;%%

%\cite{Campbell:2007ev}
\bibitem{Campbell:2007ev}
  J.~M.~Campbell, R.~K.~Ellis and G.~Zanderighi,
  %``Next-to-leading order predictions for $WW+1$ jet distributions at the
  %LHC,''
  arXiv:0710.1832 [hep-ph].
  %%CITATION = ARXIV:0710.1832;%%

\bibitem{inprep1}
   T. Binoth, J.-Ph. Guillet, S. Karg, N. Kauer, G. Sanguinetti; in preparation. 

\bibitem{inprep}
   T. Binoth, J.-Ph. Guillet, G. Heinrich, T. Reiter; in preparation. 


  \end{thebibliography}
\end{document}